\documentclass{article}

\usepackage[preprint,nonatbib]{neurips_2022}

\usepackage[utf8]{inputenc} %
\usepackage[T1]{fontenc}    %
\usepackage{hyperref}       %
\usepackage{url}            %
\usepackage{booktabs}       %
\usepackage{amsfonts}       %
\usepackage{nicefrac}       %
\usepackage{microtype}      %
\usepackage{xcolor}         %
\usepackage{algorithm,algorithmicx,algpseudocode}
\usepackage{makecell}
\usepackage{multirow}
\usepackage{float}
\usepackage{graphicx}
\usepackage{subfig}
\usepackage{amsmath}
\usepackage{wrapfig}
\usepackage[blocks]{authblk}

\title{LPCSE: Neural Speech Enhancement through Linear Predictive Coding}

\author[1]{Yang Liu}
\author[2]{Na Tang }
\author[3]{Xiaoli Chu }
\author[4]{Yang Yang }
\author[5]{Jun Wang }
\affil[1,3]{The University of Sheffield, Sheffield, United Kingdom}
\affil[2]{University of Chinese Academy of Sciences, Shanghai, China}
\affil[4]{Terminus Group, China}
\affil[5]{University College London, London, United Kingdom}
\affil[ ]{Email: \{$^{1}$yangliu,$^{3}$x.chu\}@sheffield.ac.uk, $^{2}$tangna@mail.sim.ac.cn, $^{4}$dr.yangyang@terminusgroup.com, $^{5}$jun.wang@cs.ucl.ac.uk}

\begin{document}

\maketitle
\begin{abstract}
    The increasingly stringent requirement on quality-of-experience in 5G/B5G communication systems has led to the emerging neural speech enhancement techniques, which however have been developed in isolation from the existing expert-rule based models of speech pronunciation and distortion, such as the classic Linear Predictive Coding (LPC) speech model because it is difficult to integrate the models with auto-differentiable machine learning frameworks.
    In this paper, to improve the efficiency of neural speech enhancement, we introduce an LPC-based speech enhancement (LPCSE) architecture, which leverages the strong inductive biases in the LPC speech model in conjunction with the expressive power of neural networks.
    Differentiable end-to-end learning is achieved in LPCSE via two novel blocks: a block that utilizes the expert rules to reduce the computational overhead when integrating the LPC speech model into neural networks, and a block that ensures the stability of the model and avoids exploding gradients in end-to-end training by mapping the Linear prediction coefficients to the filter poles.
    The experimental results show that LPCSE successfully restores the formants of the speeches distorted by transmission loss, and outperforms two existing neural speech enhancement methods of comparable neural network sizes in terms of the Perceptual evaluation of speech quality (PESQ) and Short-Time Objective Intelligibility (STOI) on the LJ Speech corpus.
\end{abstract}

\section{Introduction}

With the development of 5G/B5G communication systems and the rise of artificial intelligence, high-quality speech communications are required in various applications, such as  Automatic Speech Recognition (ASR), Augmented Reality (AR), and semantic communication \cite{9685250}.
However, the speech will inevitably be interfered with by the process of transmission, such as background noise, sound transmission loss (STL)\cite{Bies2017}, and multipath fading \cite{8737366}.
Particularly, a speech usually suffer from an obvious frequency-dependent STL when transmitting through objects, such as water \cite{4752682}, face masks \cite{Corey2020}, bone conduction \cite{10.1007/978-3-030-04021-5_14}, and walls \cite{Bies2017}. 
STL will distort the speech formants which are the most sonorant components in a syllable, and will detrimentally reduce the intelligibility.

Speech enhancement (SE) aims at improving the quality and intelligibility of target speech and
suppressing unwanted distortions.
Classic expert-rule based methods tried to enhance the formant-distorted speech, such as Linear Predictive Coding (LPC) \cite{MELLAHI2015545}, and Wiener filtering \cite{536932}, but their expressive powers are typically restricted.
With the development of machine learning (ML), many neural SE networks have been proposed for denoising or dereverberation \cite{8707065, HuLLXZFWZX20, YinLuoXiongZeng2020}. They exploit the well-structured spectrums or the periodic tones as the input features for SE, but the features are usually damaged due to the frequency-dependent STL, which bring challenges to existing neural SE methods.
Furthermore, most purely data-driven ML approaches are unable to extract interpretable knowledge from data and may be physically inconsistent or implausible \cite{Karniadakis_2021}. To address this challenge, some initial works are focusing on combining expert-rule based models with neural networks, such as the grey box system simulation \cite{9488228}, and the LPC-based speech vocoder \cite{8682804}.

Despite achieving impressive results by neural SE, three challenges remain under-explored. Firstly, although many expert-rule based technologies work on various types of distortions, most neural network-based efforts still focus on noise suppression or dereverberation, while how to use neural networks against the formant distortion has not been well studied.
Secondly, it is not straightforward to design a simple and compact SE network to identify and estimate the effects of STL on a speech without assuming any prior knowledge of the transmission channel.
Last but not least, most neural networks are often inefficient since they do not utilize the existing knowledge of how speech is generated or distorted, but how to use the strong inductive biases in the rules of speech pronunciation and distortion without losing the expressive power of neural networks remains an open problem.

In this paper, we propose an LPC-based single-channel SE network architecture (LPCSE), which integrates an interpretable LPC speech model into auto-differentiable neural networks and combines the strong inductive biases in expert rules with the benefits of neural networks.
Specifically, we introduce a differentiable block, which utilizes the sparsity of a matrix of Linear Prediction (LP) coefficients to overcome the challenge of large computational overhead when integrating the LPC speech model into neural networks. In addition, we propose another block that maps the LP coefficients to the poles of the LPC filter, to ensure the stability of the model and avoid exploding gradients in end-to-end learning. 
The experiments on the LJ Speech corpus demonstrate that LPCSE successfully restores the formants distorted by the STL without the need for large-scale, complex networks or models, and obtains significant gains in terms of the Perceptual evaluation of speech quality (PESQ) \cite{rec2005p} and Short-Time Objective Intelligibility (STOI) \cite{5713237} as compared with two existing neural SE methods of comparable network sizes. 
The contributions of this paper are summarized as follows:

\begin{itemize}
  \item We introduce LPCSE, an expert-rule inspired network architecture, to perform end-to-end SE for formant-distorted speeches. LPCSE enables utilizing the strong inductive biases of the LPC speech model while retaining the expressive power of SE neural networks and the benefits of end-to-end learning. 
  \item We propose two new blocks to overcome the large computational overhead when integrating the LPC speech model into neural networks and ensure the stability of the model in end-to-end learning, respectively.
\item Our ablation and comparison experiments on the LJ Speech corpus verify that the combination of the LPC speech model and neural networks in LPCSE can provide a larger gain in terms of the speech quality and intelligibility than each of them working alone. 
The LPCSE outperforms two existing neural SE methods of comparable network sizes in terms of PESQ and STOI.
\end{itemize}

\begin{figure}
    \centering
    \subfloat[The process of the LPC speech production and distortion  \label{fig:1_0}]{\includegraphics[width=0.47\textwidth]{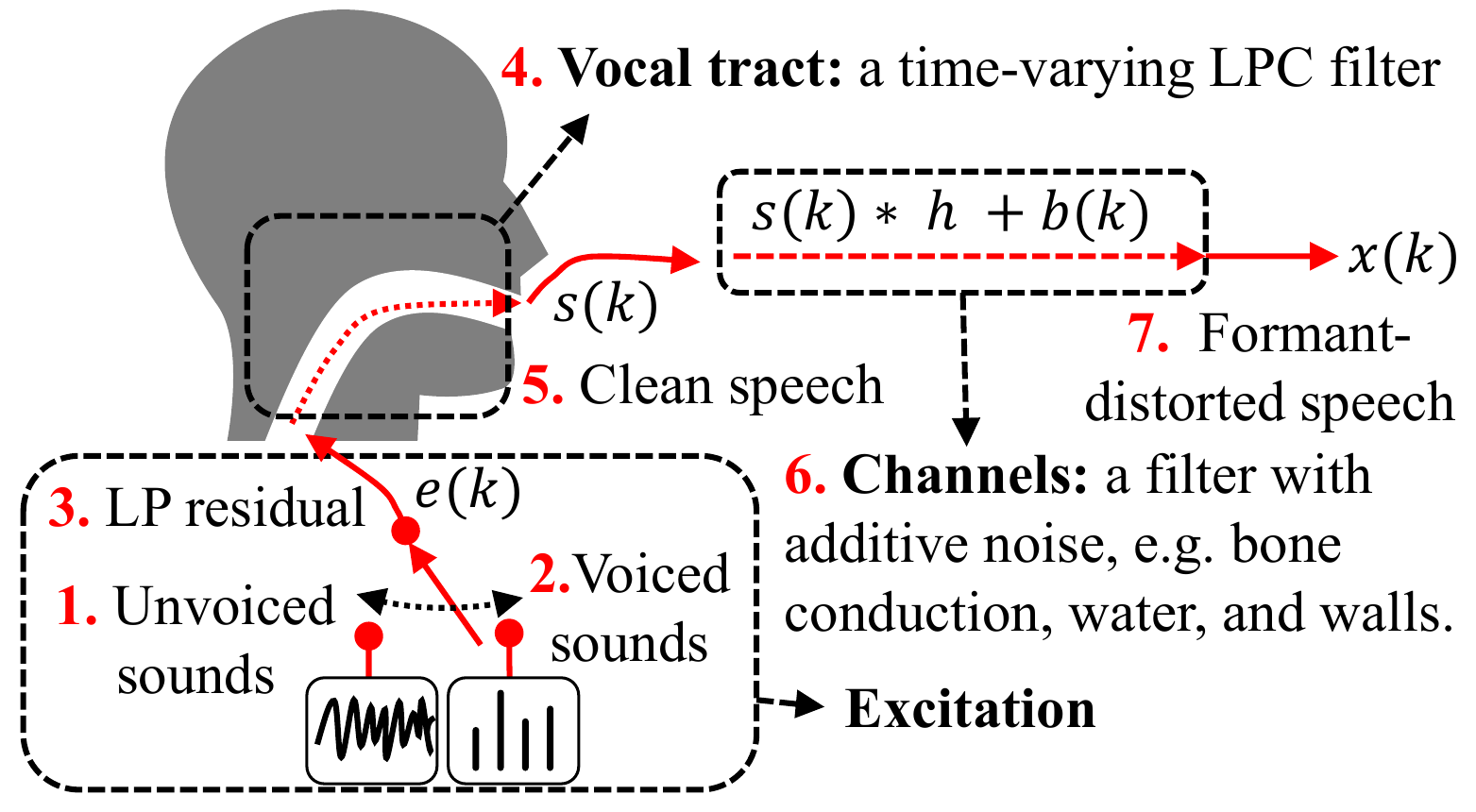}}
    \quad
    \subfloat[An example of formant distortion \label{fig:1_1}]{\includegraphics[width=0.47\textwidth]{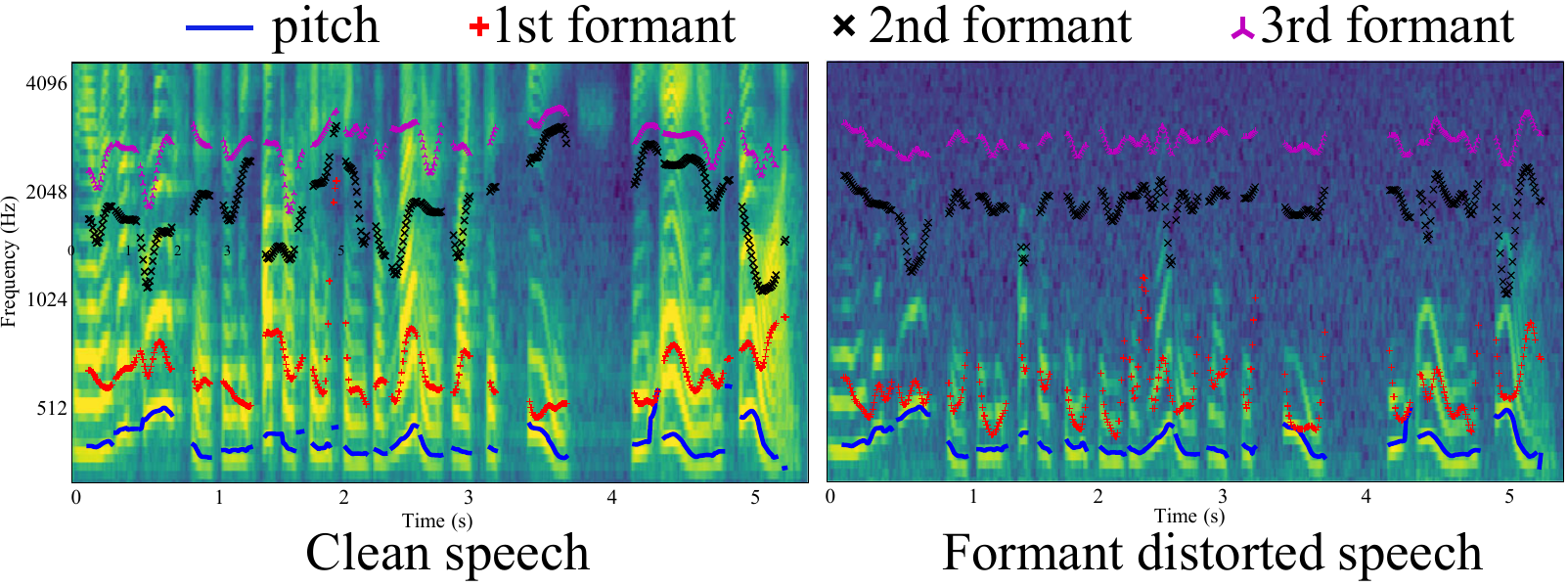}}
    \caption{The LPC speech model and formant distortion}
    \label{fig:1}
\end{figure}

\section{LPC-based Neural Speech Enhancement}
In this section, we will introduce the expert rules of the LPC and formulate the problem of LPC-based neural SE.

\subsection{Background - the Expert Rules of LPC}
Speech is the production of our phonatory systems, including lungs, larynx, vocal tract, etc., which can be well modeled by an LPC speech model \cite{1451722}.
The vocal tract response can be interpreted as an LPC filter, i.e., a time-varying all-pole linear filter with the technique of LPC. As shown in Fig. \ref{fig:1_0}, speech $ s(k) $ is produced by passing an excitation signal $ n(k) $ through the LPC filter, and can be written as follows:
\begin{equation}\label{eq1}
  s\left( k \right) =\sum_{p=1}^P{a}_{p}s\left( k-p \right) +n\left( k \right),
\end{equation}
where $ k $ denotes the time index, $ P $ is the order of the filter, $ a_p $, $p=1,...,P$, represents the LP coefficients, which hinge on the vocal tract and remain unchanged for short segments of speech, e.g., a few milliseconds.
The excitation signal $n(k)$ is related to the pronunciations of words and can be either a train of periodic impulses for voiced sounds, such as nasals and vowels, or a random noise source for unvoiced sounds.
By analyzing speech $s(k)$ and its past samples with LPC, the LP coefficients $ a_p $ and excitation $ n(k) $ can be determined from short segments of speech by minimizing the sum of squared differences between the actual and predicted speech samples.

\begin{wrapfigure}[12]{r}{0.35\textwidth}
    \centering
        \includegraphics[width=0.35\textwidth]{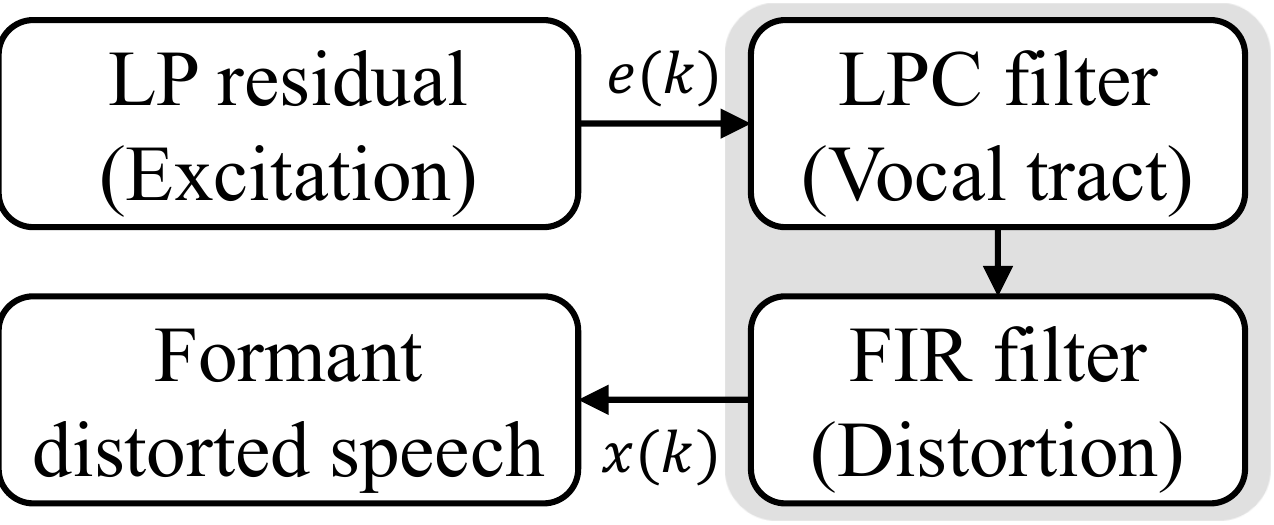}
        \caption{The formants produced by the vocal tract are disrupted by the distortion.
        }
        \label{fig:2_0}
\end{wrapfigure}

The speech observed by a reciever, denoted by  $ x(k) $, is the convolution of the speech $ s(k) $ and the channel impulse response $ h $ of the environment plus additive noise $ b(k) $, i.e., 
\begin{equation}\label{channel}
  x(k) = s\left( k \right) * h  +b\left( k \right),
\end{equation}
where $ * $ represents linear convolution.
For an environment with fixed distortion, such as speech transmission through a wall, it is usually assumed that $ h $ is time-invariant (or short-time time-invariant) and can be represented as a finite impulse response (FIR) filter \cite{Bies2017}.
To find out the relationships between formant-distorted speeches and their clean pairs, it is assumed that the excitation signal $ n(k)$ remains unchanged after the distortion \cite{10.1007/978-3-030-04021-5_14}, and the environment only affects the vocal tract response, i.e., the all-pole linear filter. Note that the limited order $ P $ of the LPC filter makes it difficult to describe the complex distortion of environments, while the performance of speech enhancement is considerably dependent on the accuracy of the filter estimation. To address this challenge, we design a Mel-spectrogram enhancement network to improve speech quality.

\subsection{Problem Formulation}
Our task is to recover $ s(k) $, given only the observed speech $ x\left( k \right) $ without assuming any priori knowledge regarding the channel impulse response $ h $.
Let $\mathcal{E}$ denote the index set of $\left| \mathcal{E} \right|$ possible scenarios.
Let ${{X}^{e}}$ be the observed speech, which is the same speech received in scenario $e\in \mathcal{E}$. The speech of ${{X}^{e=1}}$ denotes the scenario without suffering from the distortion and is used as ground truth data.
For $j = {1,\ldots ,m}$, let $X_j^e$ be the \textit{j-th} sample of $X^e$ received in $e$. ${{W}^{e}}\in {{\mathbb{R}}^{m\times m}}$ defines a directed acyclic graph (DAG) on $ m $ samples. Specifically, ${{W}^{e}}=\left[ w_1^e,\cdots ,w_j^e,...,w_m^e \right]$ is the LP-coefficient matrix of ${{X}^{e}}$ and $w_j^e$ is the LP coefficient of $X_j^e$.
There exists a vector of excitation signals $Z=[{{z}_{1}},{{z}_{2}},...,{{z}_{m}}]$ that is independent of the scenario $ e $ for all $e\in \mathcal{E}$.
Hence, $ X_j^e $ can be represented as:
\begin{equation}
  X_j^e=w_j^{eT}X^e+z_j, j = {1,\ldots ,m}. 
  \label{eq:xj}
\end{equation}
This is similar to the formulation for the structural equation model (SEM) in DAG learning problems \cite{NEURIPS2018_e347c514}. But the weighted adjacency matrix $ W $ in SEM is usually assumed to be fixed across all $e\in \mathcal{E}$, and the $ W $ in the LPC speech model is dependent on the scenario $ e $ and the excitation signals $Z$.
Furthermore, the $Z$ in SEM is treated as unknown random noise, but the excitation signals $Z$ can be decoded from the ${X^e}$ through the Levinson-Durbin recursion \cite{1451722}.

\begin{figure}
    \includegraphics[width=\textwidth]{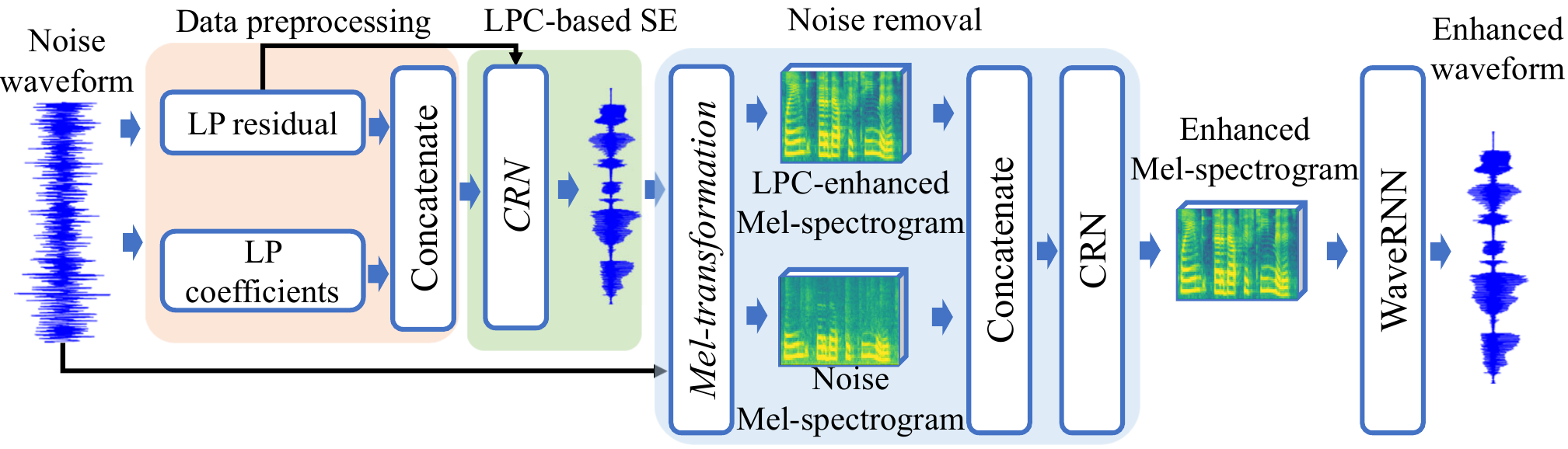}
    \caption{LPCSE architecture. \textit{CRN} denotes a convolutional-recurrent network, which will be detailed in Section \ref{NA}. \textit{Mel-transformation} converts waveforms to Mel-spectrograms, and is implemented by using the class module of \textit{torchaudio.transforms.MelSpectrogram} in Pytorch.}
    \label{fig:sys}
  \end{figure}

For brevity, we abbreviate $ e = 1 $ in the symbols of clean speech, and in the rest of this paper, $e\in \mathcal{E}$ and $e\ne 1$. Specifically, the clean speech $ X^{e=1} $ and its DAG $ W^{e=1} $ are denoted by $ X $ and $ W $, respectivly. Their disorted pairs are defined as $ X^e $ and $ W^e $.
Hence, based on Eq.(\ref{eq:xj}), the clean speech can be represented as  $  X=W^TX+Z $,
which can be reformulated as a matrix multiplication as follows:
\begin{equation}
  X=VZ,\ 
  V={{\left( I-W^T \right)}^{-1}}\in {{\mathbb{R}}^{\left( ML+1 \right)\times \left( ML+1 \right)}},
  \label{eq:xx}
\end{equation}
where $ M $ is a given number of adjacent samples that share the same LP coefficients. 
The excitation signal $ Z $ is obtained by the Levinson-Durbin recursion and the clean speech and its distorted pairs have the same $ Z $. 
We refer to $ M $ samples as a slot, and a frame has $ L $ slots.
Increasing $ L $ can reduce the number of waveform phase discontinuities and improve speech quality, but will increase computation and memory requirements.
To enhance distorted speech $ X^e $, a straightforward strategy is to use a network $\mathcal{F}$, i.e., the box of LPC-based SE in Fig. \ref{fig:sys}, to estimate the matrix $ V $ that minimizes the least square loss,
\begin{equation}\label{Problem1}
  \begin{array}{ll}
   \operatorname{min} & {{\left\| VZ-\mathcal{F}({{X}^{e}};\theta)Z \right\|}_{2}^{2}},
  \end{array}
\end{equation}
where $ \theta $ is the parameter set of the network $ \mathcal{F} $.
However, we find that although most of the speech pitch and formants are recovered by the matrix $ V $ estimated by $\mathcal{F}$, additional noises are introduced to the speech due to the limited order of the LPC speech model and the loss in the LP-coefficient matrix estimation. 
To address this challenge, another network $\mathcal{G}$, i.e., the box of noise removal in Fig. \ref{fig:sys}, is proposed to remove the noise and further enhance the speech in the time-frequency (TF) domain as follows:

\begin{equation}\label{Problem2}
  \begin{array}{ll}
   \underset{\varphi}{\min} & {{\left\| X_M-\mathcal{G}({\hat{X}^e_M};\varphi) \right\|}_{2}^{2}}                                         \\
    \text{subject to}    & \theta^{*} \in \underset{\theta}{\operatorname{argmin}}\left({{\left\| X - \mathcal{F}({{X}^{e}};\theta)Z \right\|}_{2}^{2}}\right) \\
                 & X_M = \mathcal{M}(X),\ \hat{X}^e_M = \mathcal{M}( \mathcal{F}({X^e};\theta^{*})Z;{X}^e ),          \\
  \end{array}
\end{equation}
where $ X_{M} $ denotes the Mel-spectrograms of the clean speech $ X $, and $ \hat{X}^e_M $ denotes the Mel-spectrogram of the enhanced speech $ \mathcal{F}(X^e;\theta)Z $ concatenated with the distorted speech $ X^e $.
$ \mathcal{M} $ represents the transformation from the audio waveforms to Mel-spectrogram. $ \varphi $ denotes the parameters of $\mathcal{G}$. The Mel-spectrogram transformation is differentiable and can be implemented by the standard modules in ML frameworks, such as the \textit{torchaudio} module in \textit{Pytorch} \cite{yang2021torchaudio}.

  \section{Network Architecture}\label{NA}
\begin{figure}
    \centering
    \includegraphics[width=0.65\textwidth]{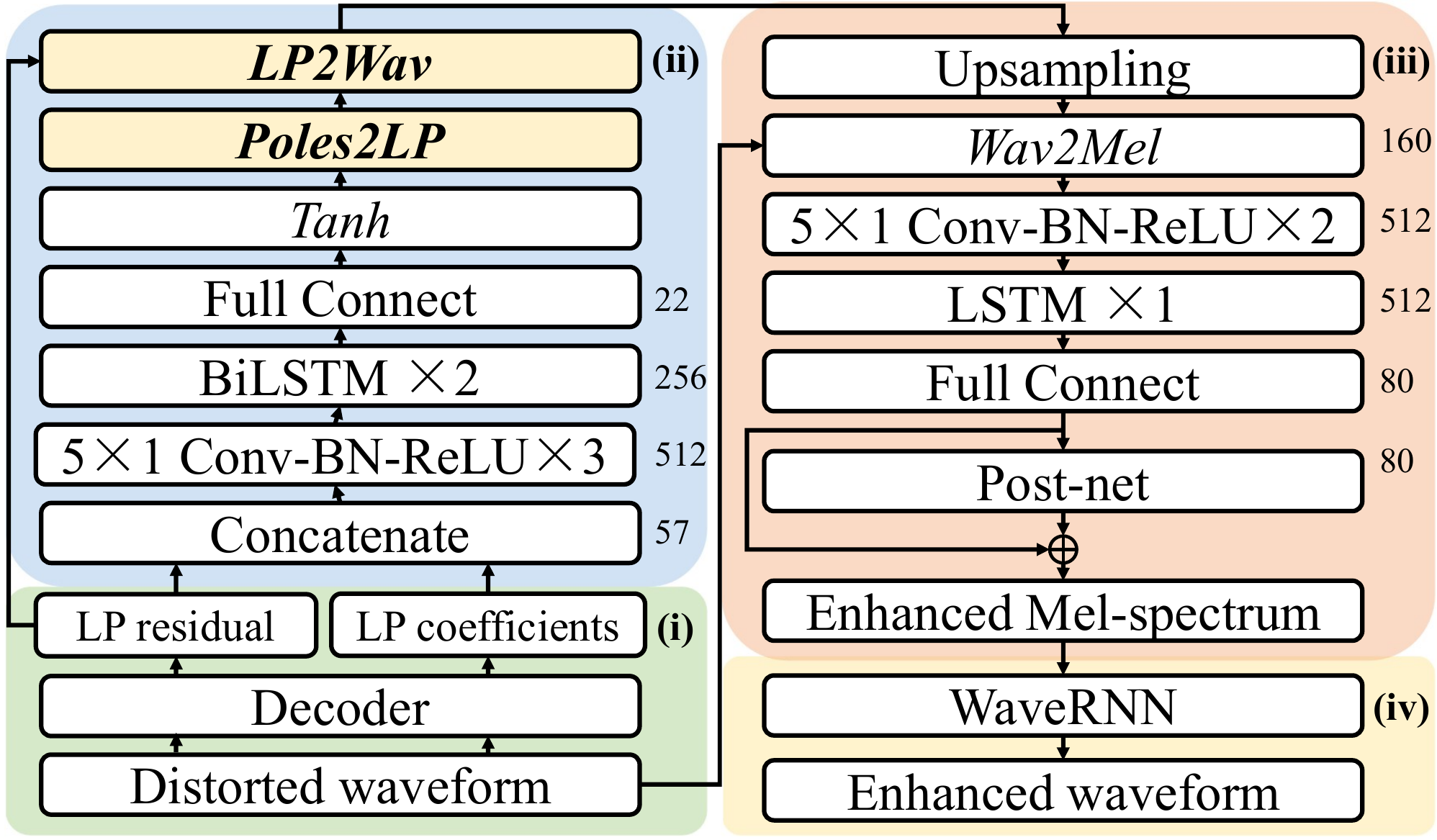}
    \caption{The detailed LPCSE architecture. \textit{LP2Wav} and \textit{Poles2LP} denote the proposed blocks. 
    \textit{Conv-BN-ReLU} and \textit{Conv-BN-Tanh} denote the convolution with batch normalization followed by \textit{ReLU} and \textit{Tanh} activation, respectively. Their kernel size is  $ 5 \times 1 $. \textit{LSTM} denotes the Long Short Term Memory network. \textit{BiLSTM} repesents bi-directional LSTM. The number on the right of each layer/block represents its cell/output dimension. }
    \label{fig:s1}
  \end{figure}

  LPCSE solves the SE problem with a cross-domain architecture, which includes four key components: a data preprocessing step, a time-domain network  $ \mathcal{F} $  for LP coefficients estimation and waveform generation,  a TF domain network  $ \mathcal{G} $ for Mel-spectrogram enhancement, and a vocoder.
  In the data preprocessing step, i.e., box (i) in Fig. \ref{fig:s1}, a distorted speech is decoded into the LP coefficients and residuals with an LPC order of $ P $ and a step size of $ M $ samples.
  The network $ \mathcal{F} $, i.e., box (ii) in Fig. \ref{fig:s1}, concatenates $ M \times L $ LP coefficients and their residuals as input. In this module, a stack of convolutional and recurrent networks with a \textit{Tanh} layer is used to extract the features from the concatenated LP coefficients and residuals, and estimate the poles of the LPC filter. Two dedicated blocks, i.e. \textit{LP2Wav} and \textit{Poles2LP}, are proposed to transfer the poles to LP coefficients and ensure the stability of the LPC speech model respectively. 
  In the network $ \mathcal{G} $, i.e., box (iii) in Fig. \ref{fig:s1}, the waveform from the \textit{LP2Wav} block is upsampled to 22 kHz by interpolating to match the sampling rate of the Mel-spectrogram vocoder and its clean speech pairs. In the \textit{Wav2Mel} block, the upsampled waveform is concatenated with the distorted speech, and transformed into two 80-channel log-Mel spectrograms corresponding to the waveforms using the 2048-point Short-Time Fourier Transform (STFT) with a \textit{Hann} window of 50 ms frame length and 12.5 ms frameshift. 
  Finally, the Mel-spectrogram is further enhanced via a convolutional-recurrent network and a \textit{Post-net} \cite{8461368} to denoise and improve speech quality. 
  In box (iv) in Fig. \ref{fig:s1}, the enhanced Mel-spectrogram is converted into an enhanced waveform through \textit{WavRNN}\cite{pmlr-v80-kalchbrenner18a}. 
  Such an architecture takes advantage of the well-structured Mel-spectrograms and the time-domain knowledge from the LPC speech model.
  \textit{LP2Wav} and \textit{Poles2LP} will be introduced in the next subsections.

  \subsection{The \textit{LP2Wav} Block: how to integrate the LPC speech model into neural networks?}\label{how1}
  To achieve end-to-end learning, the LPC speech model will be integrated into the network $ \mathcal{F} $ as a whole.
  Since $ V \in {{\mathbb{R}}^{\left( ML+1 \right)\times \left( ML+1 \right)}} $ is a dense matrix, directly predicting $ V $ from $ {{X^e}} $ is a challenge, owing mainly to the high computational overhead using standard networks to approximate $ V $ when $ V $'s size is large. In this subsection, we will address this the challenge in two steps: sparse matrix transformation and compression.
  In the \textbf{first} step, we use an inverse operation to convert the dense matrix $ V $ into a sparse matrix. Specifically, we find that the inverse of $ V $ is a sparse matrix based on the definition of the LPC speech model. Let $U=I-{{W}^{T}}={{V}^{-1}}$ be the inverse of $ V $, which is always invertible. 
  The derivative of $V$ can be obtained from $ {I}'=(UV)'=U{V}'+{U}'V=0  $ and $ {V}'=-V{U}'V $,
  where ${U}'=-({{W}^{T}}{)}'$ and $(\cdot {)}'$ is the derivative operator. $W \in {{\mathbb{R}}^{\left( ML+1 \right)\times \left( ML+1 \right)}}$ is a strictly upper triangular sparse matrix, which has 0 along its diagonal as well as the lower portion.
  Based on the derivative of $ W^T $, the gradient of $V$ is calculated by the chain rule, and the parameters of network $ \mathcal{F} $ and $ \mathcal{G} $ are updated by the backpropagation algorithm.
  In the \textbf{second} step, we compress the sparse matrix $ W $ obtained from the first step into a smaller matrix to remove redundant information and reduce computation. 
  Specifically, $ W $ is mapped to a smaller LP-coefficient matrix ${A}'$. 
  Each column of ${A}'$ contains the LP coefficients for a speech sample. Then, ${A}'$ is downsampled to $A \in {{\mathbb{R}}^{P\times L}}$ without losing information because the LPC speech model assumes that the samples in a short frame have the same LP coefficients. The downsampling factor is set to $ M $.
  We use the proposed networks to estimate the LP coefficients $ A $ of the distorted speech, instead of directly predicting $ V $, based on which much performance improvement is achieved. For example, $ M $ and $ P $ are set empirically to 46 and 11 in our experiments, i.e., $ (ML+1)^2 \gg PL $, so the network size and computational overhead are significantly reduced.

\begin{figure}
    \centering
    \includegraphics[width=0.65\textwidth]{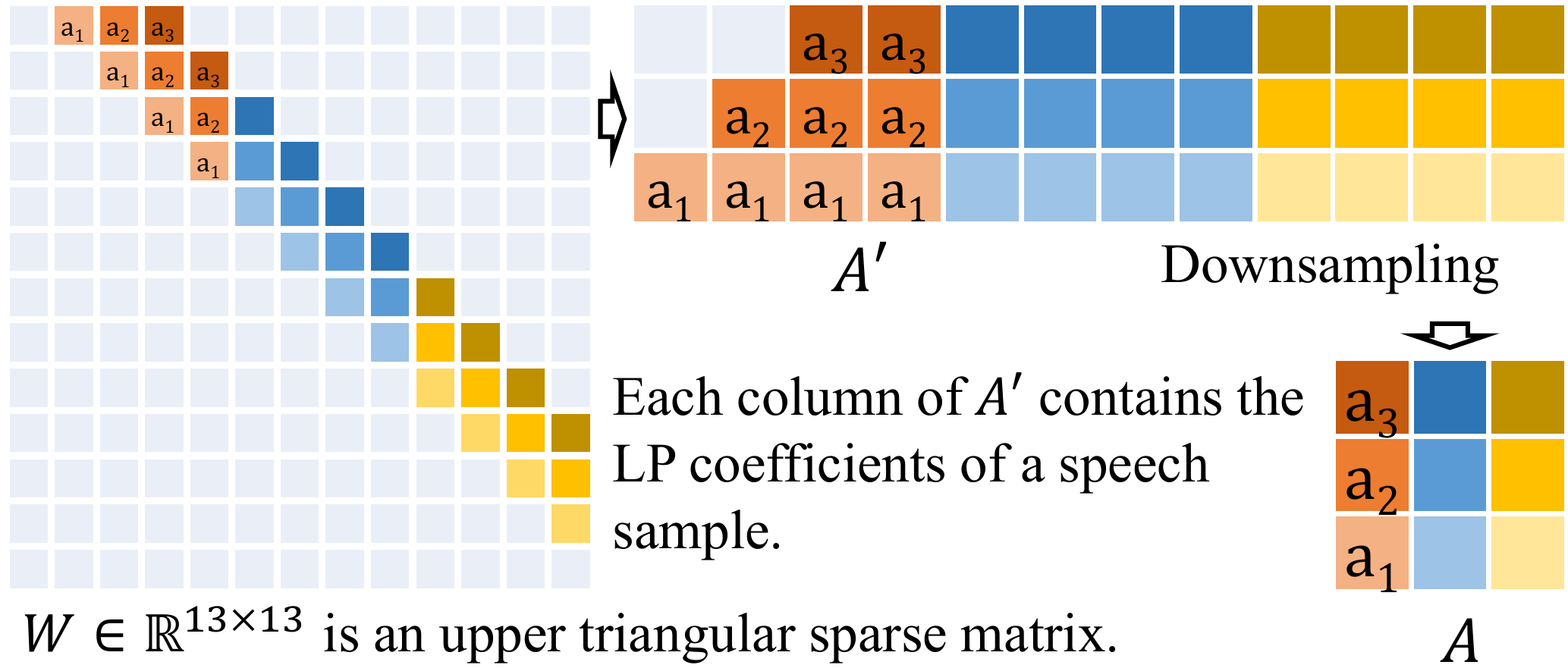}
    \caption{The expert rules reduce the computational overhead when integrating the LPC speech model into netrual networks.}
    \label{fig:3}
  \end{figure}

  To achieve the transformation from the estimated $ A $ to $ V $, we design a differentiable block, called \textit{LP2Wav}, as shown in Alg.\ref{alg:0}, where $ batch $ is the batch size.
  The \textbf{first} step in the algorithm represents the reverse process of the compression from $ {A}' $ to $ A $. The \textbf{second} step denotes the mapping from the matrix $A'$ to the sparse matrix $ W $, which is achieved by using index and in-place operations without making a copy. In the \textbf{third} step, $I-W^T$ is always invertible and its inverse can be easily calculated by using forward substitution because the index and in-place operations ensure that $I-W^T$ is an upper triangular sparse matrix with 1 along its diagonal. $ I $ is the identity tensor with the same shape as $ W $.
  To better understand the \textit{LP2Wav} block, an example illustrates the compression operation for the sparse matrix $ W $, as shown in Fig. \ref{fig:3}. For 12 speech samples and $ P = 3 $, $M$ and $L$ are set to 4 and 3, respectivly. $W\in {{\mathbb{R}}^{13\times 13}}$ is downsampled to $A\in {{\mathbb{R}}^{3\times 3}}$, without losing information.

  \begin{algorithm}
    \floatname{algorithm}{Algorithm}
    \algrenewcommand\algorithmicrequire{\textbf{Input: }}
    \algrenewcommand\algorithmicensure{\textbf{Output: }}
    \caption{The process of the \textit{LP2Wav} block}
    \label{alg:0}
    \begin{algorithmic}[1]
      \Require The estimated LP-coefficient matrix $ A(batch,L,P)$.
      \Ensure Updated $ V(batch,ML+1,ML+1) $.
      \State Obtian $A'(batch,ML,P)$ through the interpolation and concatenation to $A(batch,L,P)$.
      \State Update $ W(batch,ML+1,ML+1) $ by using the index and in-place operations to $ A'(batch,ML,P)$.
      \State $ V(batch,ML+1,ML+1) \leftarrow (I-W(batch,ML+1,ML+1)^T)^{-1} $.
    \end{algorithmic}
  \end{algorithm}

  \subsection{The \textit{Poles2LP} Block: how to guarantee the stability of the LPC speech model in neural networks?}\label{how2}
  Stability is the most fundamental of model properties.
  When using a network to estimate the paraments for an LPC speech model, the most basic requirement is that the paraments should ensure the stability of the LPC system.%
  Lacking stability will introduce infinity values into $ V $ and cause exploding gradients in the inverse operation of $I-W^T$. In this subsection, we will introduce how to guarantee stability using standard operations and blocks in neural networks.

  Based on Eq.(\ref{eq1}), the system transfer function of the LPC speech model is an all-pole linear filter, and can be rewritten as follows by using the \textit{z}-transform,
  \begin{equation}
    \begin{aligned}
    l(z)=&\frac{K}{1-{{a}_{1}}{{z}^{-1}}-{{a}_{2}}{{z}^{-2}}\cdots -{{a}_{P}}{{z}^{-P}}} 
    =\frac{K{{z}^{P}}}{{{z}^{P}}-{{a}_{1}}{{z}^{P-1}}-{{a}_{2}}{{z}^{P-2}}\cdots -{{a}_{P}}},
    \end{aligned}
  \end{equation}
  where the denominator polynomial coefficients $\{{{a}_{1}},{{a}_{2}},...,{{a}_{P}}\}$ are real numbers. $ K $ is a given system gain, which is set to 1 in our experiments. In order to guarantee the stability, the locations of $l(z)$'s poles must lie inside or on a unit circle on the complex plane. By the fundamental theorem of algebra, $l(z)$ has $P$ poles, which occur as complex conjugate pairs or real numbers. Let $\{{{r}_{1}},{{r}_{2}},...,{{r}_{P}}\}$ be the poles of $l(z)$. 
  A straightforward idea is to use a supervised learning model to estimate $\{{{r}_{1}},{{r}_{2}},...,{{r}_{P}}\}$ with the constraint of $\left| {{r}_{i}} \right|\le 1$ for $i=1,2,...,P$ which will guarantee the stability, instead of directly estimating $\{{{a}_{1}},{{a}_{2}},...,{{a}_{P}}\}$. 
  However, after obtaining the poles in this way, how to enable the transformation from the pole $ r_i $ to the coefficient $ a_i $ in neural networks has not been well studied.

\begin{figure}
    \centering
    \includegraphics[width=0.65\textwidth]{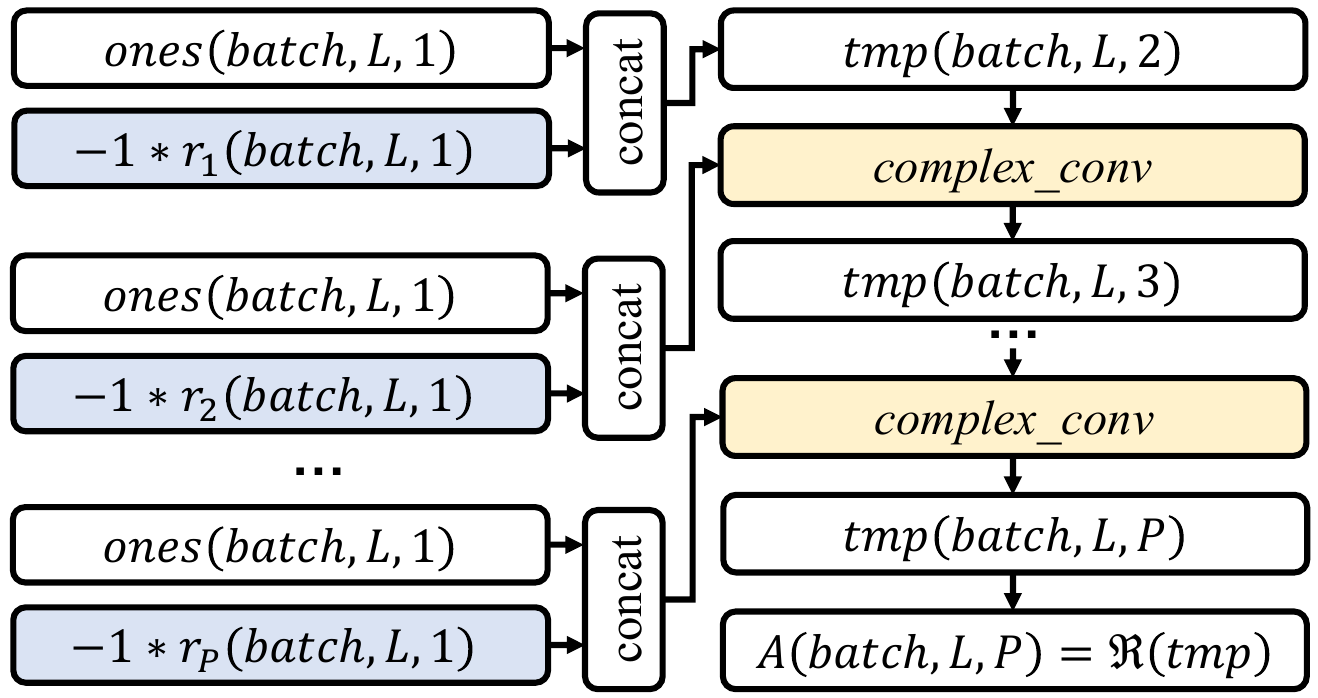}
    \caption{The flowchart of the proposed \textit{Poles2LP} block.}
    \label{fig:4}
\end{figure}

  To bridge the gap, we propose a differentiable block for the transformation in the network $ \mathcal{F} $, called \textit{Poles2LP}, which maps the poles of $l(z)$ to the LP coefficients in $ A $ using only standard operators and blocks in ML, as shown in Fig. \ref{fig:4}. \textbf{Firstly}, $l(z)$ can be rewritten as:
  \begin{equation}
    l(z)=\frac{Kz^P}{\left( z-{{r}_{1}} \right)\left( z-{{r}_{2}} \right)\cdots \left( z-{{r}_{P}} \right)},
  \end{equation}
  where $\{{{r}_{1}},{{r}_{2}},...,{{r}_{P}}\}$ denote the poles of $l(z)$, and their real and imaginary parts are constructed by the output of the $ Tanh $ layer in the network $ \mathcal{F} $.
  \textbf{Secondly}, since polynomial multiplication is equivalent to convolution, the LP coefficients are calculated through the iterative convolution between the vector $[1,-{{r}_{i}}]$ and convolution results, $i=1,2,..., P$. 
 A tensor of $ones(batch, L,1) $ is concatenated with $ -{{r}_{i}}(batch, L,1) $ in the last dimension. 
  \textbf{Thirdly}, a complex convolution block \cite{HuLLXZFWZX20} calculates the iterative convolution between the concatenated tensor and previous convolution results.
  In the complex convolution, let $Re_1$ ,$Re_2$ be the real part of two complex convolution block inputs, and $Im_1$, $Im_2$ be the imaginary part of the inputs as follows
  \begin{equation}
    \begin{aligned}
       & Re=conv\left( Re_1,Re_2 \right) -conv\left( Im_1,Im_2 \right) \\
       & Im=conv\left( Re_1,Im_2 \right) +conv\left( Im_1,Re_2 \right), \\
    \end{aligned}
  \end{equation}
  where $Re$ and $Im$ denotes the real and imaginary part of the complex convolution block output, and $conv()$ denotes the standard convolution operator for real numbers, which is implemented by using existing convolution operators, such as \textit{torch.nn.functional.conv1d} function in Pytorch.
  \textbf{Finally}, after the iteration for the $ P $ poles, the real part of the final convocation results is the downsampled LP-coefficient matrix $ A(batch,L,P)$, which contains $\{{{a}_{1}},{{a}_{2}},...,{{a}_{P}}\}$ in its columns.
  A flowchart is shown in Fig.\ref{fig:4} to describe the process of the \textit{Poles2LP} block, where $ \Re $ is the real part symbol, and $ tmp $ is the convolution result. $ P $ is a small value, e.g. 11 in our experiment. $ complex\_conv $ represents the complex convolution.

  \subsection{Loss Functions}\label{loss}
  Our goal is to solve the following optimization problem:
  \begin{equation}\label{4.6}
    \underset{\theta,\varphi}{\mathop{\min }}\,L={{L}_{Mel}}+\mu {{L}_{Wave}}+\lambda {{L}_{LP}},
  \end{equation}
  where $ \mu $ and $ \lambda $ are hyperparameters chosen for balance. $L_{Mel}$ is a Mel reconstruction loss, $L_{Wave}$ is a waveform reconstruction loss, and $ L_{LP} $ is a LP-coefficient loss.
  
  \textbf{Mel reconstruction loss.}
  We refer to the output of the full connect layer in $ \mathcal{G} $ as the coarse Mel-spectrogram, denoted by ${{\hat{X}}^{e}}_{M'}$. 
  We refer to the sum of the coarse Mel-spectrogram and the output of the \textit{Post-net} as the fine-grained Mel-spectrogram, denoted by ${{\hat{X}}^{e}}_{M}$ , and  ${{\hat{X}}^{e}}_{M} =\mathcal{G}(\mathcal{M}( \mathcal{F}(\tilde{X}^e;\theta)Z;\tilde{X}^e);\varphi) $, respectively.
  We use the sum of the mean squared error (MSE) between the target Mel-spectrogram $ X_M $ and the coarse and fine-grained Mel-spectrograms as the Mel reconstruction loss, i.e., $ {{L}_{Mel}}=\mathbb{E}\left[ \left\| {{{\hat{X}}}^{e}}_{M}-X_M \right\|_{2}^{2} \right] + \mathbb{E}\left[ \left\| {{{\hat{X}}}^{e}}_{M'}-X_M \right\|_{2}^{2} \right] $.

  \textbf{Waveform reconstruction loss.}
  We will refer to the upsampled waveform in the network $ \mathcal{G} $ as the coarse-grained speech, denoted by $ \hat{X}_{c}^e $, i.e., $ \hat{X}_{c}^e = \mathcal{F}(\tilde{X}^e;\theta)Z $.
  We use the MSE between the target $ X $ and upsampled waveform $ \hat{X}_{c}^e $ as the waveform reconstruction loss, i.e., $ {{L}_{Wave}}=\mathbb{E}\left[ \left\| {\hat{X}}_{c}^e-{{X}} \right\|_{2}^{2} \right] $.
  
  \textbf{LP-coefficient loss.} 
  We use an LP-coefficient loss to the loss function, denoted by $ {{L}_{LP}} $ in Eq.(\ref{4.6}), which represents the expectation of the complex-valued \textit{L1} norm between the estimated LP coefficients and the target LP coefficients, i.e., $ L_{LP}=\mathbb{E}\left[ \lVert \hat{A}^e-A \rVert _{1}^{2} \right] $,
  where $ \hat{A}^e $ and $ A $ are the estimated and target LP coefficients, respectively. $ \hat{A}^e $ is the input of the \textit{Poles2LP} block.
  $ A $ is obtained from the clean speech pairs of $ X^e $ in the training dataset.
  Note that $ \hat{A}^e $ is a complex tensor, and $ A $ is a real tensor. During training, the imaginary part of $ \hat{A}^e $ will approach 0, and the real part will be close to $ A $.
\section{Experiments}

\subsection{Dataset and implementation details}

\textbf{Dataset.}  In our experiments, the LJSpeech corpus \cite{ljspeech17} is used to evaluate the proposed LPCSE. LJSpeech is a public data set and contains 13,100 audio clips of a single speaker reading passages with a sampling rate of 22 kHz. 
To obtain a clean dataset, we remove the silences in the dataset using the WebRTC voice activity detection (VAD)\cite{WebRTC}, since the silences do not correspond to any vocal changes.
To construct distorted-clean speech pairs, we let the clean speech pass through an FIR filter, which is used to simulate the channel impulse response $ h $ of a 5 cm thick concrete wall. The FIR filter is designed according to the Sharp's equation \cite{Bies2017}, which is widely used to estimate the STL of walls in the noise control of buildings. Then, we mix the filtered speech with pink noise at 3 Signal-to-noise ratio (SNR) levels from -3 to 3 dB. 

\begin{figure}
    \centering
    \includegraphics[width=0.65\textwidth]{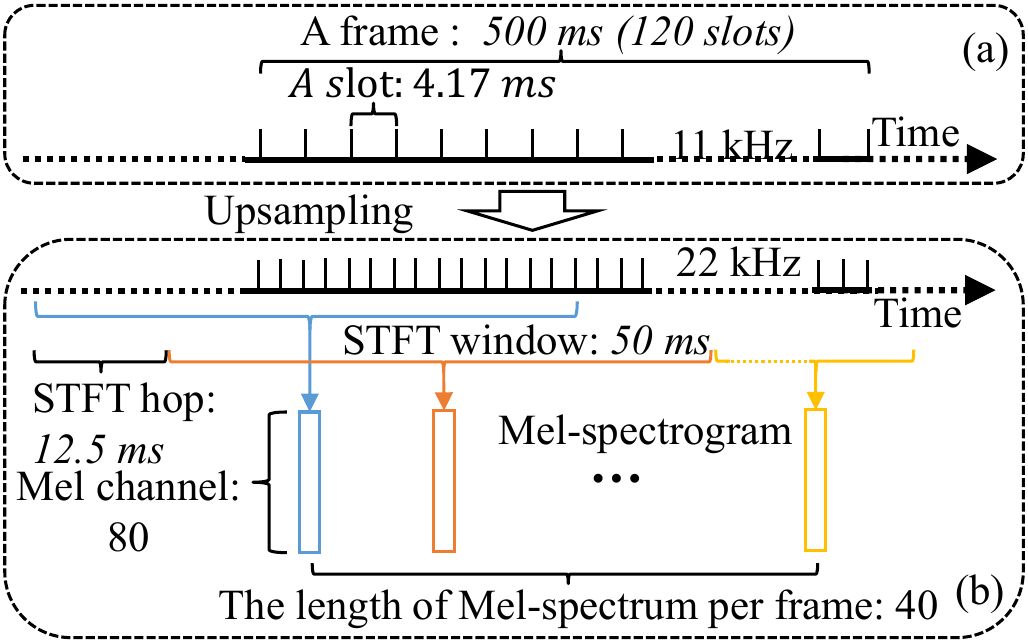}
    \caption{The settings for the frames and slots in our experiments. (a) In the network $ \mathcal{F} $, a frame containts 120 slots, and each slot has 46 samples for the sampling rate of 11 kHz. (b) After upsamlping to 22 kHz in the network $\mathcal{G} $, the waveforms are transformed to Mel-spectrograms by employing the STFT.}
    \label{fig:5}
\end{figure}

\textbf{Implementation Details.}
The distorted-clean speech pairs are divided into training and test sets by 9:1. We use the Perceptual evaluation of speech quality (PESQ) \cite{rec2005p} and Short-Time Objective Intelligibility (STOI) \cite{5713237} as the evaluation metric to measure the quality and intelligibility of speech, respectively. 
$ \mu \in[0, \infty) $ and $ \lambda \in[0, \infty) $ in the loss function balance the contribution of the network $ \mathcal{F} $, $ \mathcal{G} $ and the LPC speech model. The hyperparameters are set to $ \mu =1 $, $ \lambda = 0.3 $ in our experiments. We train our model on 2 NVIDIA 2080Ti GPUs with a batch size of 16 frames on each GPUs.  Each frame contains $L$ slots, and each slot has $M$ samples, and they are set to  $L = 120$, and $ M =46 $, as shown in Fig. \ref{fig:5}. We use the ADAM optimizer with a fixed learning rate of 0.001.

\subsection{Ablation Study}\label{Ablation_Study}
\begin{figure}
    \centering
    \includegraphics[width=0.65\textwidth]{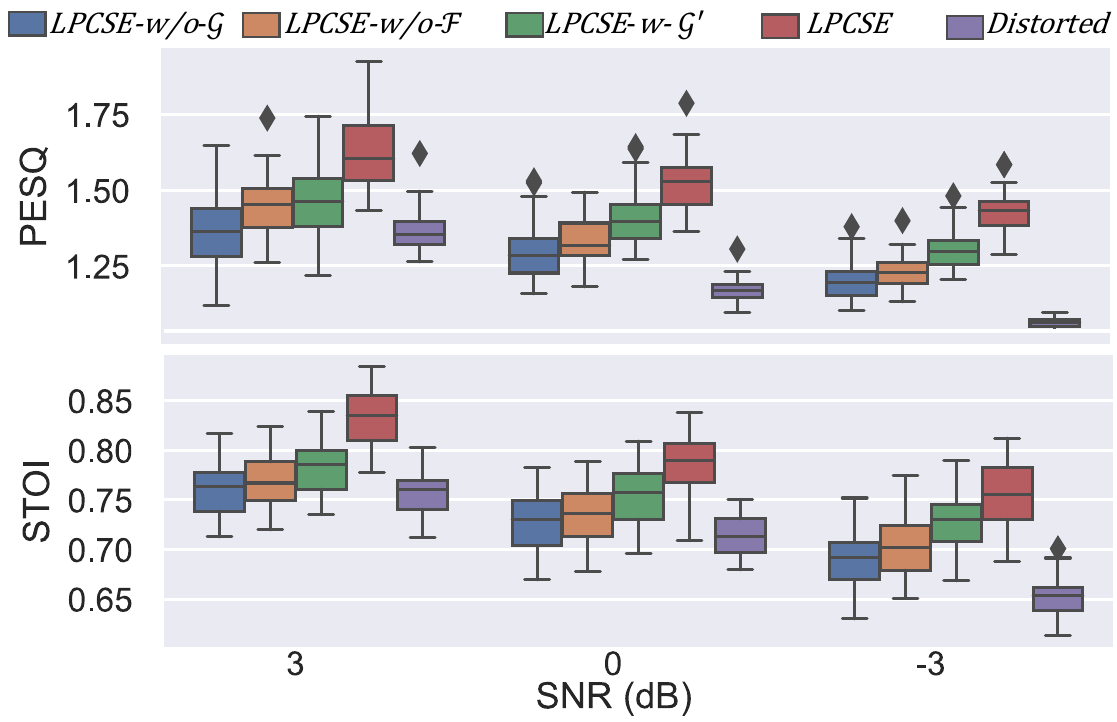}
    \caption{PESQ and STOI comparisons of the ablation study.}
    \label{fig:9}
\end{figure}

An ablation study is conducted to identify the essential components of the proposed architecture. The ablation results are shown in Fig. \ref{fig:9}.

\textbf{LPCSE and Distorted:}
among these methods, LPCSE represents our full architecture, which includes the proposed network $ \mathcal{F} $ and $ \mathcal{G} $ . \textbf{Distorted} denotes the formant distorted speech without enhancement.
A significant gain of 0.25-0.38 and 0.07-0.11 from LPCSE is observed in median PESQ and STOI when compared with \textbf{Distorted}. The results show that the proposed network can improve speech quality and intelligibility in different scenarios. LPCSE significantly outperforms other methods in the ablation study.

\textbf{LPCSE-w/o-$ \mathcal{G} $:}
LPCSE-w/o-$ \mathcal{G} $ shows the performance of only using the network $ \mathcal{F} $ and removing the network $ \mathcal{G} $ from the LPCSE architecture, except for the \textit{LP2Wav} and upsampling layer in the network $ \mathcal{G} $. The network $ \mathcal{F} $ estimates the clean LP coefficients based on the excitation signal and distorted LP coefficients, and the LP coefficients is fed into the \textit{LP2Wav} block. The output of LPCSE-w/o-$ \mathcal{G} $ is the upsampled waveform from the output of the \textit{LP2Wav} block, instead of the Mel-spectrogram from the network $ \mathcal{G} $.
The results in Fig.\ref{fig:9} show that, compared with the LPCSE architecture, LPCSE-w/o-$ \mathcal{G} $ only provides a small gain of 0.02-0.14 and 0-0.03 on PESQ and STOI, respectively. This is expected because LPCSE-w/o-$ \mathcal{G} $ only uses a simple traditional method, i.e., the LPC speech model, to synthesize speech, which is difficult to mitigate the noise in generated speech. The aim of LPCSE-w/o-$ \mathcal{G} $, including the network $ \mathcal{F} $, \textit{LP2Wav} block and upsampling layer, is to recover the distorted high-frequency components and provide coarse Mel-spectrograms, based on which the rest of the network $ \mathcal{G} $ will generate fine-gain Mel-spectrograms.

\textbf{LPCSE-w/o-$\mathcal{F}$ and LPCSE-w-$\mathcal{G}'$:}
LPCSE-w/o-$\mathcal{F}$ and LPCSE-w-$\mathcal{G}'$ are the two settings that remove the network $ \mathcal{F} $ from the LPCSE architecture, while the network $ \mathcal{G} $ 
 is preserved. The difference is that the LPCSE-w/o-$\mathcal{F}$ uses the same network $ \mathcal{G} $ as LPCSE, and LPCSE-w-$\mathcal{G}'$ uses an enhanced network $ \mathcal{G}' $, which has twice the size of $ \mathcal{G}$. We see that LPCSE-w/o-$\mathcal{F}$ could provide a gain of 0.1-0.16 and 0.01-0.05 on median PESQ and STOI, respectively.
LPCSE-w-$\mathcal{G}'$ has a larger capacity for speech enhancement than LPCSE-w/o-$\mathcal{F}$ but compared with LPCSE-w/o-$\mathcal{F}$, it only introduces a small increase of 0.01-0.08 and 0.02-0.04 on the median PESQ and STOI, respectively. The gains of both settings drop significantly as the distortion and noise become strong, i.e., from  -3 dB to 3 dB, because the network $ \mathcal{G} $ is insufficient to enhance the speech alone. Comparison between LPCSE-w/o-$\mathcal{F}$ and LPCSE shows that our architecture could provide 0.15-0.22 and 0.05-0.08 gain on median PESQ and STOI in different scenarios. This proves that, compared with the architectures which only use the network $ \mathcal{G} $, LPCSE has a more stable performance as the distortion and noise increase.

In summary, the results of the ablation study suggest that both the network $ \mathcal{F} $ and $ \mathcal{G} $ are crucial to the LPCSE architecture. Jointly integrating the networks in our design significantly outperforms the methods where any of them is applied separately since the combination of them is a more effective way to improve speech quality and intelligibility than only using the network $ \mathcal{F} $ or $ \mathcal{G} $.

  \begin{table}
    \centering
    \caption{Comparison of different methods}
    \setlength{\tabcolsep}{3.1pt}
    \scalebox{1}{
    \label{tab:3}
    \begin{tabular}{*{16}{c}}
      \toprule
      Methods       & \multicolumn{3}{c}{PESQ} & \multicolumn{3}{c}{STOI}  &\multirow{2}{*}{\makecell[c]{F/B GPU \\ Time (ms)} } & \multirow{2}{*}{\makecell[c]{Params \\(million)}}\\
      \cmidrule(r){2-4} \cmidrule(r){5-7}  
      SNR (dB)      & \textit{3}       & \textit{0}       & \textit{-3}       & \textit{3}       & \textit{0}       & \textit{-3}  \\
      \midrule
  
      Distorted     & 1.36           & 1.17           & 1.06           & 0.76           & 0.71           & 0.65  &   ---  & ---    \\
      \makecell[c]{EHNET}  & 1.23           & 1.21           & 1.23           & 0.80           & 0.77           & 0.73   & 28/84 & 17.6      \\
      \makecell[c]{PHASEN} & 1.12           & 1.17           & 1.24           & 0.44           & 0.43           & 0.45    & 12/46 & 7.36   \\
      LPCSE       & \textbf{1.62}      & \textbf{1.53}      & \textbf{1.41}      & \textbf{0.83}      & \textbf{0.79}      & \textbf{ 0.76} & 105/164  & 15.5\\
      \bottomrule
    \end{tabular}
    }
  \end{table}

\subsection{System Comparison}
We compare LPCSE with two existing neural SE networks, EHNET \cite{8462155} and PHASEN \cite{YinLuoXiongZeng2020}, in terms of PESQ and STOI, and the number of trainable parameters (Params), as shown in Table \ref{tab:3}.
To compare the complexity of the methods, we also measure the forward and backward (F/B) GPU time for an input speech of 7 seconds.
EHNET is a convolutional-recurrent method worked in the TF domain, which only uses the Mel-spectrograms of the distorted speech as input. 
PHASEN is a cross-domain method and captures both amplitude and phase-related information of distorted speeches. 
The results show that LPCSE significantly outperforms ENNET and PHASEN in terms of PESQ and STOI because the LPC speech model uses the mathematical model with stability constants that make the search space of the parameters tractable. 
However, in exchange, LPCSE requires a slightly larger computation of 102 and 164 ms GPU time in the F/B propagation, respectively, and the computation comes mainly from the inverse operation in the \textit{LP2Wav} block. 
An example of the LPCSE's result is shown in Fig. \ref{fig:mel1}. From the figure, we can see that the poles of enhanced speech are close to that of the clean speech, and the first formant difference between enhanced and clean speech has a significant peak at 0, which means the formants are correctly restored.

  \subsection{Visualizations}
\begin{figure}
    \centering
    \subfloat[The distorted, enhanced and clean Mel-spectrograms of a speech.\label{fig:mel1.noisy}]{\includegraphics[width=0.8\textwidth]{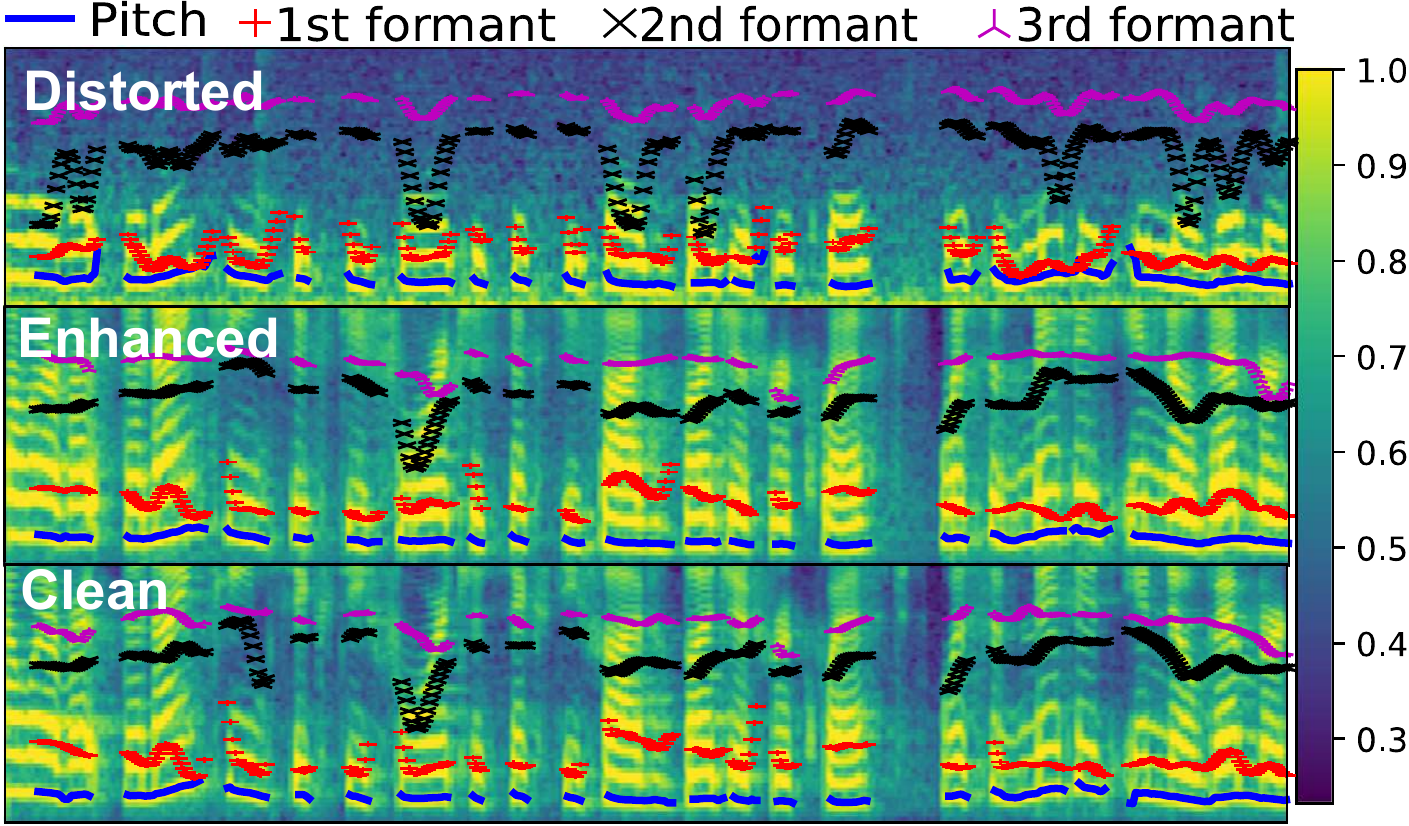}}
  \\
    \subfloat[The positions of the poles on the complex plane. The poles of enhanced speech are close to that of clean speech, which indicates that the LP coefficients are corrected after SE. \label{fig:poles}]{\includegraphics[width=0.47\textwidth]{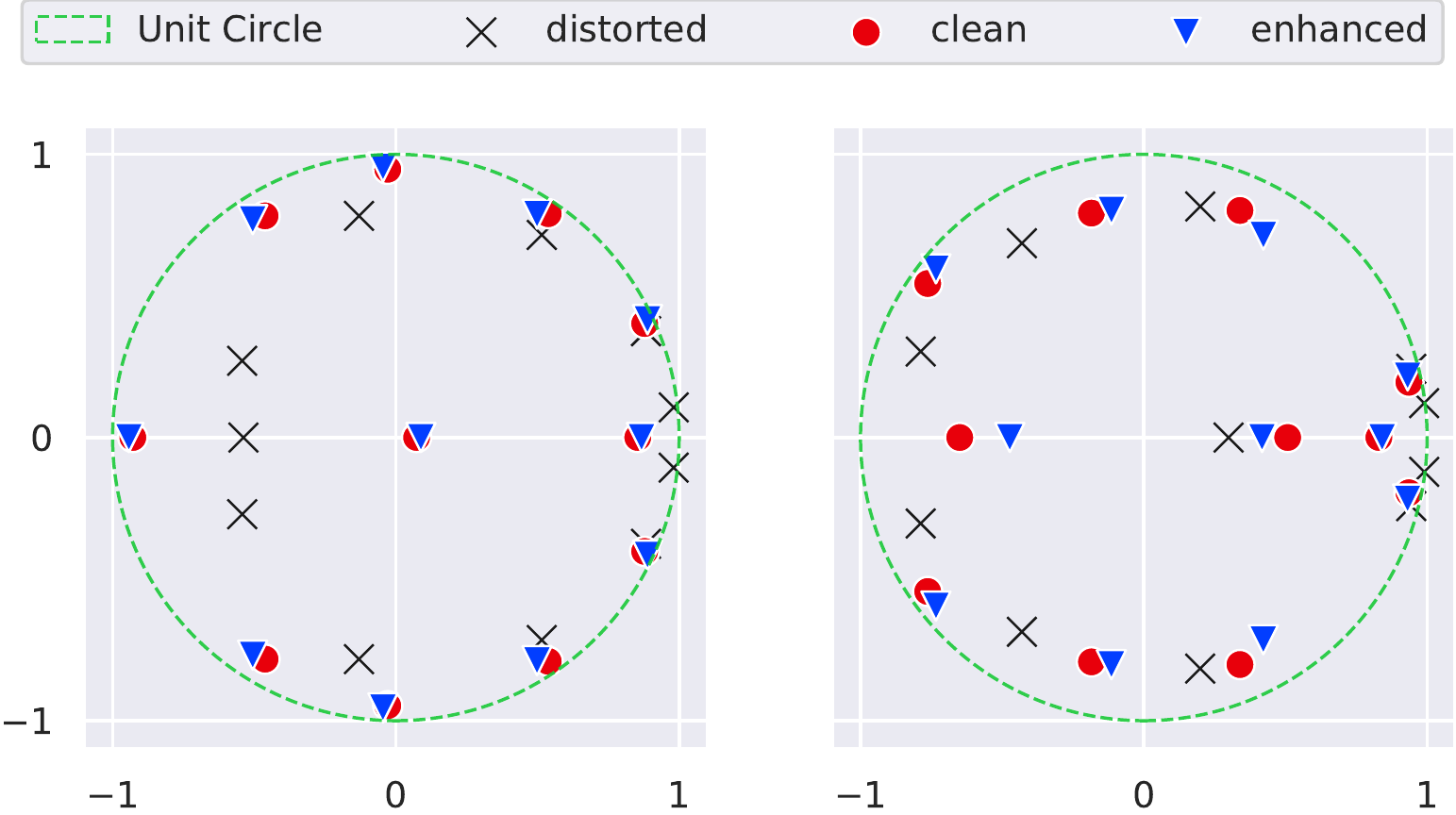}}
    \quad
    \subfloat[The formant differences between the distorted/enhanced and clean speech. The peaks of the blue blocks are close to 0, which indicates the distorted formants have been recovered after speech enhancement.\label{fig:F1F2}]{\includegraphics[width=0.47\textwidth]{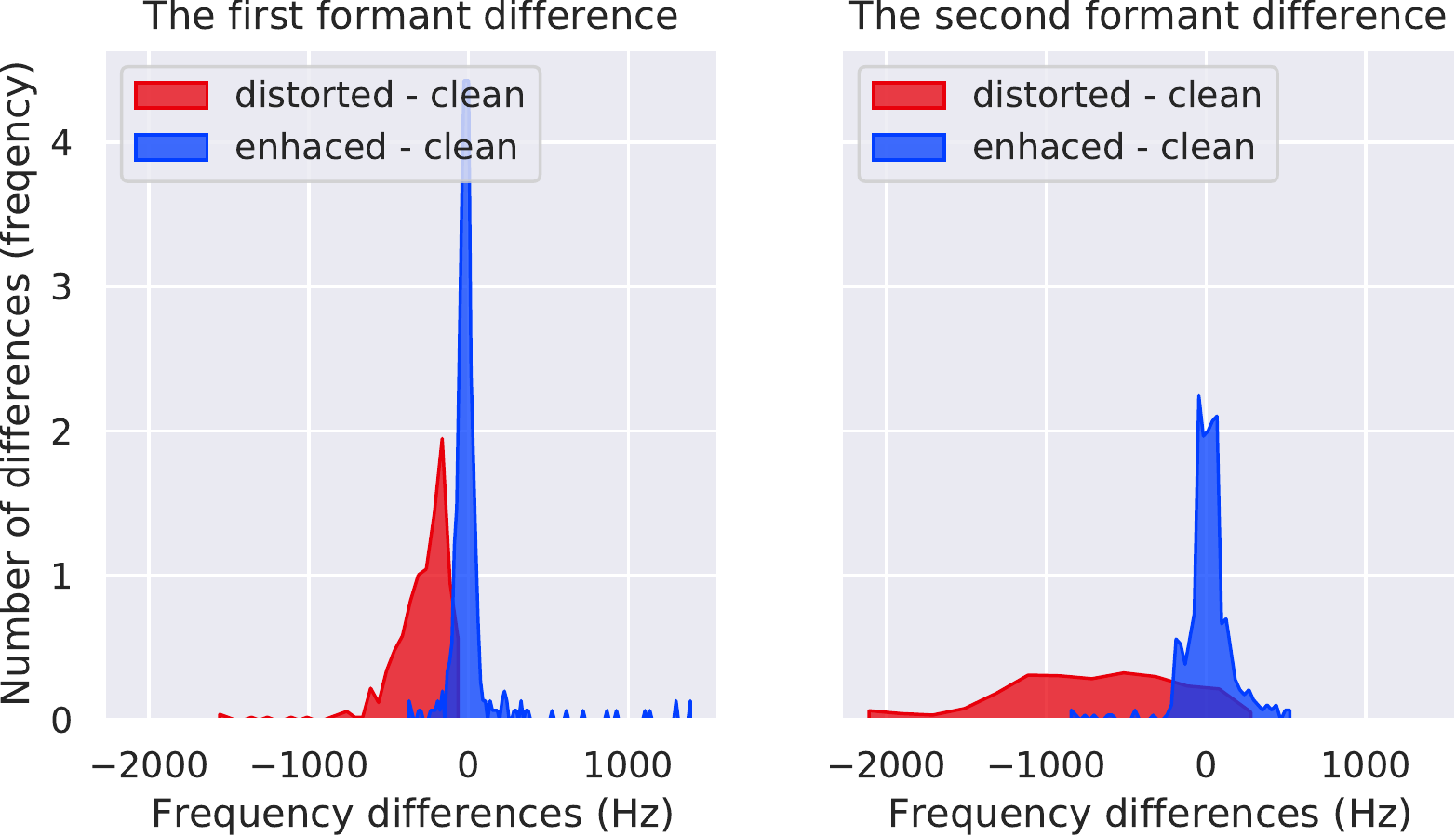}}
    \caption{Comparisons of Mel-spectrograms, poles and formants a formant distorted speech, and its LPCSE-enhanced, and clean pairs from the test dataset.}
    \label{fig:mel1}
  \end{figure}

  To have a better understanding of the LPC speech model in LPCSE, we do a case study by visualizing the enhanced speech.
  Fig.\ref{fig:mel1}(a) shows the Mel-spectrograms, pitch, and formants of a formant distorted speech, and its LPCSE-enhanced, and clean pairs from the test dataset. The formants are calculated by Burg's algorithm for linear prediction coefficients \cite{Andersen1974OnTC}. The pitches are computed by an acoustic periodicity detection algorithm based on an accurate autocorrelation method \cite{Boersma93accurateshort-term}.
  Compared with the clean speech, the high-frequency components of the formant distorted speech are damped by noise, and the pitch is less affected due to the low-frequency components maintained. After the SE through LPCSE, the high-frequency components are recovered in the enhanced speech which restores the formants.
  The poles of the LPC speech model in two slots of the speeches are shown in Fig.\ref{fig:mel1}(b). Compared with the distorted speech, the poles of the enhanced speech are more close to that of the clean speech, which indicates that the LP coefficients of the enhanced speech are corrected by LPCSE. 
To intuitively show the performance of formant recovery, the first and second formant differences between the distorted/enhanced and clean speech are shown as the red and blue blocks in Fig.\ref{fig:mel1}(c), respectively.
  For the formant differences between distorted and clean speech, we can see that the first formant difference has an obvious peak of about -200 Hz. The second formant difference has a flat distribution along the x-axis, which indicates the second formant suffers from more attenuation than the first one when passing through the walls because it has higher frequencies from about 1 kHz to 2 kHz.
For the formant difference between enhanced and clean speech, we can see that the differences between the formants become small, and the mean frequency differences are close to 0, which indicates most of the formants are restored by LPCSE, and the formants of the enhanced speech are close to that of the clean speech.

  \section{Related Works}

  This section reviews the typical single-channel SE methods, in which noisy speech can be enhanced either in the TF domain or time domain. 
  In TF domain methods, the speech is transformed into the frequency domain to obtain the amplitude and phase of its spectrogram, e.g. by using the STFT. Early TF domain methods try to find a multiplicative mask for the amplitude to separate target speech from background interference. The enhanced time-domain speech can be produced by applying an inverse transformation, e.g. the inverse STFT, on the enhanced amplitude and the noisy phase. For example, RDL-Net \cite{Nikzad2020} fuses both ResNets and densely connected CNNs to generate enhanced amplitude, and uses the noisy phase and an inverse STFT-based framework to produce enhanced speech.
  As the importance of the phase has been realized in SE, phase-aware and complex-valued methods have been proposed to simultaneously estimate the magnitude and phase of the target speech \cite{HuLLXZFWZX20,7364200, YinLuoXiongZeng2020, Liu2021, HASANNEZHAD20221}.
  However, the phase prediction is still challenging in SE, because the phase spectrogram is highly unstructured and will significantly deviate with interferences \cite{takahashi18_interspeech}.
  To avoid phase prediction in TF domain methods, neural speech vocoders, e.g. WaveNet \cite{vandenoord16_ssw}, WaveRNN \cite{pmlr-v80-kalchbrenner18a}, and MelGAN \cite{NEURIPS2019_6804c9bc}, have been used to synthesis speech waveform by conditioning the vocoder on enhanced spectrogram \cite{8462417,du20c_interspeech}.

  Time-domain methods try to avoid the phase prediction problem by directly enhancing the raw waveform.
  The direct regression methods \cite{8331910, Stoller2018} use convolutional neural networks on the waveform to learn the regression function from the input speech to the target speech.
  SEGAN \cite{Pascual2017} directly enhances the input waveform in the time domain by generative adversarial networks (GANs). Conv-TasNet \cite{8707065} uses a learnable encoder-decoder in the time domain as an alternative to the STFT for speech separation tasks. 
  Wave-U-Net \cite{Stoller2018} uses the U-Net framework to combine features at multiple time scales in the time domain for audio source separation tasks. The time-domain methods can not utilize the well-structured patterns on TF spectrograms, which will potentially lead to a performance loss. To fully leverage the information in the time and TF domain, HiFi-GAN \cite{su20b_interspeech} uses multi-domain adversarial discriminators to train the WaveNet network for denoising and dereverberation. 
  These methods achieve remarkable performance on their goal by avoiding the problems in the TF domain.
  However, the time-domain methods are usually complex and cumbersome compared with the TF domain methods, because the processing of waveform is more difficult than that of the well-structured TF spectrogram.
  To leverage the TF spectrogram in the time domain, TFT-Net \cite{ijcai2020-528} proposed a cross-domain framework that learns a multiplicative mask for the spectrogram and computes loss for the decoded waveform in the time domain.
  Compared with the time-domain and TF domain methods, our work proposes an innovative expert-rule inspired network architecture, which could utilize the time and TF domain features while keeping a simple and compact network architecture.

  To reduce the network complexity and avoid learning from scratch, some works have tried to estimate the parameters for expert-rule based speech models, instead of directly predicting speech waveforms or spectrums.
  In particular, a convolutional recurrent network \cite{9449307} predicts the speech amplitude spectrum for a Wiener filter gain function to minimize the mean square error in the process of noise-suppressing. A DNN-based SE algorithm \cite{9180931} predicts clean LPC coefficients for a given distorted speech, which are used by a Kalman filter to recover the clean speech spectrogram. 
  However, the methods fall short of achieving end-to-end learning. Further, the expert rules are imperfect, and the gap between practical conditions and ideal assumptions in the models, e.g. the assumption of the best filtering effect, and stationary Gaussian distribution of noise, will limit the quality of the generated speech.
  To address this challenge, some initial works try to compensate for the imperfect by integrating the expert-rule based models into neural networks.
  In particular, LPCNet \cite{8682804} and LP-WaveNet \cite{9306362} have shown the advantages of LPC-based networks in improving the efficiency of speech synthesis.
  DDSP \cite{engel2020ddsp} uses a harmonic plus noise model to realize audio synthesis, which directly integrates an autoencoder architecture with classic signal processing parameters, e.g. fundamental frequency and loudness. 
  However, LPCNet requires clean LP coefficients as input, which is difficult to obtain in SE.
  DDSP has a good performance in music synthesis, but it's difficult for the harmonic plus noise model to generate high-fidelity speech as harmonic contents in speech are more complex than that in music.

\section{Conclusion}
In this article, we have investigated an LPC-based SE architecture, named LPCSE, which aims to enhance the speeches distorted by the frequency-dependent STL through the combination of LPC and neural networks.
Our experiment results have shown that the proposed LPCSE can correct and recover the formants of the distorted speeches. According to our ablation study, the LPC speech model embedded in LPCSE plays a key role in improving the speech quality and intelligibility, thereby demonstrating the effectiveness of the LPC in neural SE.
In the comparison with two existing neural SE methods of comparable network sizes, LPCSE has larger improvements in terms of PESQ and STOI, which indicates that LPCSE opens a new path toward simpler and better SE systems with the help of expert rules.

\bibliographystyle{plain}
\bibliography{sample-base}

\end{document}